\documentclass[english]{revtex4}
\usepackage[T1]{fontenc}
\usepackage{amsmath}
\usepackage{amssymb}
\usepackage{graphicx}

\makeatletter
\@ifundefined{textcolor}{}
{%
 \definecolor{BLACK}{gray}{0}
 \definecolor{WHITE}{gray}{1}
 \definecolor{RED}{rgb}{1,0,0}
 \definecolor{GREEN}{rgb}{0,1,0}
 \definecolor{BLUE}{rgb}{0,0,1}
 \definecolor{CYAN}{cmyk}{1,0,0,0}
 \definecolor{MAGENTA}{cmyk}{0,1,0,0}
 \definecolor{YELLOW}{cmyk}{0,0,1,0}
}

\makeatother

\usepackage{babel}
\begin{document}
\title{Local Short-Time Acceleration induced Spectral Line Broadening and
Possible Implications in Cosmology}
\author{M.J.Luo}
\address{Department of Physics, Jiangsu University, Zhenjiang 212013, People's
Republic of China}
\email{mjluo@ujs.edu.cn}

\begin{abstract}
The paper proposes an acceleration effect that a local short-time
acceleration produces an additional broadening to spectral line, while
the central value of the line remains unaffected. The effect can be
considered as a local and non-uniform generalization of Unruh effect.
Although the acceleration-induced line broadening effect is too small
to be measured in ordinary lab setup, it may offer us a key concept
to gain a simple and unified perspective on the cosmic acceleration
and the radial acceleration discrepancy of rotating galaxies. We find
that the measurement of the acceleration of the cosmic expansion by
fitting the distance-redshift relation is essentially the measurement
of the line or redshift broadening, and the cosmic acceleration induced
line broadening also plays a crucial role in the acceleration discrepancy
at the outskirt of rotating galaxies. Possible predictions of the
effect are also discussed.
\end{abstract}
\maketitle

\section{Introduction}

Spectral line emitted from astronomical body is an important observation
object in astrophysics and cosmology at long distance scale. Many
important information of the distant astronomical bodies and the universe
can be extracted from it. For example, the expansion of the universe
or the Hubble's law are indicated by the observation of the redshift
of spectral lines of distant stars. In general, since one can measure
the velocity of a moving body at cosmic distance by the Doppler shift
of spectral lines it emits, how can one measure the acceleration of
the moving body from its spectral lines? Especially, how the spectral
method can be used to understand the acceleration expansion of the
universe, and the acceleration discrepancy \citep{Stacy2016Radial,2017ApJ...836..152L,Li:2018tdo}
at the outskirt of galaxies? 

To our knowledge, the Unruh effect provides us an acceleration effect
to spectrum, but it applies to a uniform and long time acceleration
in flat spacetime background. The expansion acceleration at any local
point in spacetime are divergent in their directions and hence non-uniform,
and the radial accelerations of a rotating galaxy are along the radial
directions and hence also non-uniform. What is the local and non-uniform
analogous version of the Unruh effect? On the other hand, as we know,
the measurements of the luminosity of distant spectral lines of type
Ia supernovae (SNIa) as function of their redshifts (so-called distance-redshift
relation) have led to the interpretation that the expansion of the
universe is presently accelerated and therefore the energy density
of the universe is seem presently dominated by a mysterious ``dark
energy'' with ``negative pressure''. So we could ask, what do we
really measure in these spectral lines of SNIa? Is there any underlying
relation between the (cosmic) acceleration and the spectral lines,
does the (cosmic) acceleration have any unknown effect on the spectral
lines, so that one can extract information about the cosmic acceleration
from the spectral lines measurements? In addition to the mysterious
expansion acceleration of the universe, there exists another discrepancy
or anomaly in acceleration measurements at the outskirt of galaxies
which are also based on spectral lines observations. The acceleration
discrepancy \citep{Stacy2016Radial,2017ApJ...836..152L,Li:2018tdo}
reveals the discrepancy between the measured radial acceleration of
rotating tracers (21 cm spectral lines) at the outskirt of a galaxy
and the predicted acceleration given by the observed mass of the galaxy
according to the Newtonian gravity. It leds to an important aspect
of the ``dark matter'' problem that the galaxies seem to be surrounded
by unobserved ``missing matter'' as extra gravitational mass source.
Similarly, we could also ask, does the rotational radial acceleration
as well as the cosmic acceleration have any unknown effects on the
spectral lines emitted from the distant rotational tracers of the
galaxy? Is there any underlying relation between the acceleration
discrepancy of galaxy and the acceleration expansion of the universe?
In this paper, we propose that a new acceleration-induced quantum
effect on local spectral lines may be a key to these questions. The
spectral line (of SNIa) broadening effect will be shown as a main
modification to the distance-redshift relation of the universe at
the second-order which is an indicator of the cosmic acceleration.
And the effect is also one of the important modifications to the Doppler
broadening of spectral lines (21cm-lines) which are indicators of
the rotational velocity and related radial acceleration at the outskirt
of galaxies. 

The paper is divided into two parts. The first part focuses on the
acceleration-induced quantum effect, especially for the spectral lines.
Past works on the quantum effect of acceleration or non-inertial motion
covers a broad ranges, from the well known Unruh thermal effect \citep{Unruh:1976db}
and acceleration-induced transparency \citep{Soda:2021sql} to the
non-inertial contribution to the geometric phase \citep{Martin-Martinez:2010gnz,Arya:2022lay}.
Closer to the present effect is various studies of accelerated free-falling
wave-pocket via solving the Schrodinger equation in a linear gravitational
potential \citep{Wadati2000The,Vandegrift2000Accelerating,Herdegen:2001tt,2006Quantum,2012Weak,Nauenberg2016Einstein,2017Does,2022Free,2022On},
which shows spatially spreading of the wave-pocket. Another relevant
work is the non-inertial correction to the Lamb shift \citep{Arya:2023okh}.
There are also various results concerning the relation between entanglement
and non-inertial motion \citep{Alsing:2003es,Alsing:2006cj,Bruschi:2011ug,Adesso:2012gw,Friis:2012tb,Toros:2019ptw},
for instance, a rotation induced entanglement \citep{Restuccia:2019chf,Toros:2022tpq}.
There are also different works investigating various properties of
non-inertial quantum system, for the sake of searching for appropriate
non-inertial quantum signatures in laboratory settings \citep{Rogers:1988zz,Chen:1998kp,Lynch:2019hmk,Lynch:2022nls}.
In the second part we discuss the possible relations between the present
effect and the cosmic acceleration and the radial acceleration discrepancy
of galaxies. There are a large number of proposed literature on each
of the two unsolved puzzles, we only mention some of them that are
highly relevant to our work. An alternative to the dark matter is
the so called MOdified Newtonian Dynamics (MOND) (see e.g. \citep{Famaey:2011kh,Milgrom:2014usa}),
in which the acceleration is a key notion, that is the law of Newtonian
gravity is strongly modified when the acceleration is lower than a
critical acceleration. However, the underlying physics of MOND, the
role of acceleration and the critical acceleration in it is yet unclear.
The situation leads to various speculative works. Milgrom \citep{Milgrom:1998sy}
suggested that MOND is originated from a Unruh-like vacuum thermal
effect of the deSitter spacetime. Verlinde \citep{Verlinde:2016toy}
treated the modified gravity as an entropic force in the thermal deSitter
vacuum. Pikhitsa \citep{Pikhitsa:2010nd}, Klinkhamer and Kopp \citep{Klinkhamer:2011un}
considered that the critical acceleration of MOND is related to the
minimum Unruh temperature of the deSitter spacetime. Frankly speaking,
the speculative relation between the acceleration-induced (or equivalently,
the deSitter background induced) Unruh-like thermal effect and the
modified gravity is lack of a solid physical foundation and far from
clear, if such a relation more or less exists. To understand the physical
foundation of MOND is also an important motivation of the paper.

The paper is organized as follows. In section II, the standard Unruh
effect induced Planckian spectrum is re-derived from the method of
coordinate transformation to an acceleration frame. In section III,
the long-time Fourier transformation in the section II is generalized
to a short-time windowed Fourier transformation, i.e. Gabor transformation,
a local short-time accelerated broadening spectrum is derived. In
section IV-A, we calculate the spectral line broadening due to the
acceleration of the cosmic expansion, and discuss the modification
to the distance-redshift. In section IV-B, the contribution of the
cosmic acceleration to the acceleration discrepancy of a rotating
galaxy is discussed based on the effect of acceleration-induced line
broadening effect. In section V, we summarize our results.

\section{Uniform Acceleration for Long Time: The Unruh Effect}

The well-known Unruh effect \citep{Unruh:1976db} claims that a uniformly
accelerated frame moving in flat spacetime carries a Planckian spectrum
with temperature $T_{unruh}=\frac{a}{2\pi}$ in which $a$ is the
acceleration of the frame w.r.t. a rest frame observer who sees the
thermal spectrum. Numerous derivations and discussions (see e.g. \citep{Crispino:2007eb}
and references therein) of the effect have appeared in literature
over the 50 years based on QFT in the Rindler metric with uniform
acceleration. In order to generalize the effect to a non-uniform and
local version, a more appropriate derivation is by using the Coordinate
Transformation Approach \citep{2004AmJPh..72.1524A}. 

Considering a monochromatic plane wave with frequency $\omega_{k}$
propagating along (e.g. $x$-direction) in a rest frame $K$. Obviously,
the spectrum of the plane wave in the rest frame is a delta function
\begin{equation}
\rho_{res}(\omega)=\delta(\omega-\omega_{k}),\label{eq:delta spectra}
\end{equation}
i.e. an infinitely narrow spectral line.

Now we perform a coordinate transform from the rest frame $K$ to
an accelerated frame $K^{\prime}$ with uniform acceleration $a$
(without lose of generality along the propagating $x$-direction).
It can be an acceleration $+|a|$ or deceleration $-|a|$ depends
on its sign. We consider the initial velocity of the accelerated frame
is zero, so the Lorentz transformation gives the velocity of the frame
at time $t$ 
\begin{equation}
v(t)=\frac{at^{\prime}}{\sqrt{1+\left(at^{\prime}\right)^{2}}}=\frac{\sinh\left(at\right)}{\sqrt{1+\sinh^{2}\left(at\right)}}=\tanh\left(at\right),
\end{equation}
in which we have used the relation $t^{\prime}=\frac{1}{a}\sinh\left(at\right)$
of the times between the accelerated and rest frames. Then the frequency
under the acceleration is 
\begin{equation}
\omega_{k}\rightarrow\omega_{k}^{\prime}(t)=\frac{\omega_{k}\left[1+\tanh\left(at\right)\right]}{\sqrt{1-\tanh^{2}\left(at\right)}}=\omega_{k}e^{at}.\label{eq:accelerated frequency}
\end{equation}

The time dependent frequency gives a time dependent phase $\varphi(t)$
to the plane wave. If we consider the acceleration is for long time,
i.e. $at$ can be large, the accelerated plane wave is then 
\begin{equation}
\psi(t)=\exp\left[i\varphi(t)\right]=\exp\left[-i\int_{-\infty}^{t}\omega_{k}^{\prime}(\hat{t})d\hat{t}\right]=\exp\left(-i\omega_{k}\frac{1}{a}e^{at}\right).\label{eq:wave function}
\end{equation}
Its (infinitely long-time) Fourier transformation yields
\begin{equation}
\tilde{\psi}(\omega)=\frac{1}{\sqrt{2\pi}}\int_{-\infty}^{\infty}dt\exp\left(-i\omega_{k}\frac{1}{a}e^{at}\right)e^{i\omega t}=\frac{1}{\sqrt{2\pi}}\frac{1}{a}\Gamma\left(\frac{i\omega}{a}\right)\left(\frac{-\omega_{k}}{a}\right)^{-i\frac{\omega}{a}}e^{-\frac{\pi\omega}{2a}},\quad(t\rightarrow\infty).\label{eq:spectra wave}
\end{equation}
Since $\left|\Gamma\left(\frac{i\omega}{a}\right)\right|^{2}=\Gamma\left(\frac{i\omega}{a}\right)\Gamma\left(-\frac{i\omega}{a}\right)=\frac{\pi a}{\omega}\sinh\left(\frac{\pi\omega}{a}\right)$,
finally, the spectrum of the accelerated plane wave can be obtained
\begin{equation}
\rho_{acc}(\omega)=|\tilde{\psi}(\omega)|^{2}=\frac{1}{a\omega}\frac{1}{e^{\frac{2\pi\omega}{a}}-1},
\end{equation}
which is nothing but the standard Planckian thermal spectrum with
the Unruh temperature $T_{unruh}=\frac{a}{2\pi}$.

The time dependent phase $\varphi(t)$ in eq.(\ref{eq:wave function})
can also be obtained by substituting coordinates of the accelerated
frame $K^{\prime}$ (Rindler coordinates) $x^{\prime}=\frac{1}{a}\cosh(at)$
and $t^{\prime}=\frac{1}{a}\sinh(at)$ into $-\varphi(t)\equiv k_{x}x^{\prime}+\omega_{k}t^{\prime}=\omega_{k}\frac{1}{a}e^{at}$
by taking the dispersion $\omega_{k}=k_{x}$.

We can see in this derivation that a uniform acceleration with long
duration time makes the spectral line $\rho_{res}(\omega)=\delta(\omega-\omega_{k})$
completely spread to a continuous Planckian spectrum $\rho_{acc}(\omega)$.
Since the Planckian spectrum has a maximal entropy, the information
of the frequency $\omega_{k}$ in the unaccelerated spectral line
is completely lost during the long time acceleration. As a consequence,
any spectra (no matter lines or bands) will be blurred to be a completely
thermalized Planckian spectrum under a long time uniform acceleration. 

\section{Non-Uniformly Accelerated Line in a Short-Time Window}

When the spectral line (\ref{eq:delta spectra}) undergoes a non-uniform
acceleration $a(x,t)$ (along the wave propagating $x$-direction
as the uniform case), locally it indeed feels like a uniform instantaneous
acceleration $a=a(x,t)$, but it must be in a sufficient short-time
interval. If the acceleration time is sufficiently shorter than the
typical thermalized time of the accelerated frame $K^{\prime}$, i.e.
$t\ll\frac{1}{T}=\frac{1}{a}$, certainly it will not be expected
still a Unruh-type effect and a thermalized spectrum.

In this section, we will generalize the previous derivation to a non-uniform
version in a short acceleration time. The first change to the previous
Unruh effect derivation is that in a short acceleration time interval
eq.(\ref{eq:accelerated frequency}) becomes the standard Doppler
shift
\begin{equation}
\omega_{k}\rightarrow\omega_{k}^{\prime}(t)\approx\omega_{k}\left(1+at\right),\quad(at\ll1),\label{eq:short time doppler shift}
\end{equation}
where $a$ is the ``instantaneous'' acceleration around $t$. So
in this case, eq.(\ref{eq:wave function}) becomes a short-time phase
shifted wave function
\begin{equation}
\psi(t)\approx\exp\left[-i\omega_{k}\left(1+at\right)t\right],\quad(at\ll1).\label{eq:short time wave func}
\end{equation}

Since the Doppler shift of the wave vector is $k_{x}\rightarrow k_{x}^{\prime}(t)\approx k_{x}(1+at)$,
and the displacement during the short-time is given by $\Delta x\sim O(at^{2})$,
so the phase shift $k_{x}at\cdot\Delta x$ of the wave vector is at
higher order $O(t^{3})$ than the exponent of eq.(\ref{eq:short time wave func})
which is of order $O(t^{2})$. The $O(t^{3})$ spatially spreading
effect has been shown in the Airy wavefunction (spatial) solution
in solving the Schrodinger equation in an accelerated free-falling
motion \citep{Wadati2000The,Vandegrift2000Accelerating,Herdegen:2001tt,2006Quantum,2012Weak,Nauenberg2016Einstein,2017Does,2022Free,2022On}.
Therefore, eq.(\ref{eq:short time wave func}) as the temporal part
of the wavefunction is the leading order result.

At this moment, it is purely a coordinate transformation of the wavefunction,
and no other extra phase shift involves. Different from some of the
papers \citep{Eliezer1977The,Herdegen:2001tt,2022Free}, which try
to keep the equivalence principle at the quantum level, it is argued
that the acceleration can be eliminated by a coordinate transformation
of the Schrodinger equation from an inertial frame to a free-falling
accelerated frame, but with an extra local $U(1)$ phase shift factor
in the wavefunction. However, there is no evidence for the validity
of the Schrodinger equation in the accelerated falling frame, and
there is certainly also no evidence, at this level, for a unification
of the gravitation (the coordinate transformation of the wavefunction)
and the electromagnetism (the extra local $U(1)$ phase factor of
the wavefunction). Therefore, in our treatment, the accelerated wavefunction
eq.(\ref{eq:short time wave func}) is purely a coordinate transformation
without any $U(1)$ phase shift factor, as the treatment of eq.(\ref{eq:wave function}).

The second change goes to the transformation to its spectral-domain.
Different from the previous infinity long-time Fourier transformation,
here we only focus on its local (Doppler) shifted spectrum during
the short acceleration time, or more precisely during a single period
of an ``instantaneous'' wave, rather than during the whole time-domain.
Such time-localized spectrum can be achieved by windowing the wave
function and then taking the Fourier transformation. A natural and
typical choice of the time-window function $W(t-t_{0},\sigma_{t})$
located round the initial accelerating time $t_{0}$ (without lose
of generality we have set $t_{0}=0$) is a Gaussian function, and
the size of the time-window is typically chosen to be a single period
of the wave (a shortest duration for a full wave), i.e. $\sqrt{\sigma_{t}}=1/\omega_{k}$,
i.e.
\begin{equation}
W(t-t_{0},\sigma_{t}=\omega_{k}^{-2})=\sqrt{\left|\omega_{k}\right|}e^{-\frac{1}{2}\omega_{k}^{2}\left(t-t_{0}\right)^{2}}.\label{eq:time-window}
\end{equation}
So the single period time-localized wave function in its spectral-domain
is
\begin{equation}
\tilde{\psi}(\omega,t\sim t_{0}=0)=\frac{1}{\sqrt{2\pi}}\int_{-\infty}^{\infty}dt\exp\left[-i\omega_{k}\left(1+at\right)t\right]\left(\sqrt{\left|\omega_{k}\right|}e^{-\frac{1}{2}\omega_{k}^{2}t^{2}}\right)e^{i\omega t},\quad(at\ll1),
\end{equation}
which is a generalization of eq.(\ref{eq:spectra wave}). This is
the windowed Fourier transformation, which is a standard technique
for time-frequency localization, often called the Gabor transformation
(or more general a wavelet transformation) \citep{1992Ten}.

Finally, the spectrum of the time localized wave within the short
acceleration time is given by
\begin{equation}
\rho_{sho-acc}(\omega)=\left|\tilde{\psi}(\omega,t\sim0)\right|^{2}=\frac{1}{\sqrt{\omega_{k}^{2}+\left(2a\right)^{2}}}\exp\left[-\frac{\left(\omega-\omega_{k}\right)^{2}}{\omega_{k}^{2}+\left(2a\right)^{2}}\right]\equiv\frac{1}{\sqrt{2\sigma_{\omega}}}\exp\left[-\frac{\left(\omega-\omega_{k}\right)^{2}}{2\sigma_{\omega}}\right].
\end{equation}

We can see that the spectral is Gaussian (it is not surprise since
the leading accelerated wavefunction is at order $O(t^{2})$ and time-window
function is also Gaussian), and the central value $\omega_{k}$ of
it remains unchanged, it coincides with the peak of the unaccelerated
delta spectral line $\rho_{res}=\delta(\omega-\omega_{k})$. Since
the central frequency is unaffected while the phase is short-time
locally shifted, so the short-time locally accelerated wave can interfere
with the original wave, leading to acceleration or gravitation interference
type of effects \citep{Colella:1975dq,1999Measurement}. 

Importantly, the spectral line is Gaussian broadened, and the variance
\begin{equation}
\sigma_{\omega}=\frac{1}{2}\left[\omega_{k}^{2}+\left(2a\right)^{2}\right]\label{eq:full width}
\end{equation}
of the broadened line has two parts. The first part comes from the
inverse of the size of the time-domain $\sigma_{t}=1/\omega_{k}^{2}$,
which is intrinsic and follows the standard uncertainty principle
of quantum mechanics. The second part is induced by the additional
contribution from the acceleration $a^{2}$ during the short-time,

\begin{equation}
\delta\sigma_{\omega}\equiv\sigma_{\omega}-\frac{1}{2\sigma_{t}}=2a^{2}.\label{eq:a^2}
\end{equation}

Several points about the result are needed to be emphasized. (i) First,
the additional line broadening $\delta\sigma_{\omega}$ must also
be interpreted as an extra quantum uncertainty or fluctuation or smearing
of the spectrum, besides the intrinsic part $1/\sigma_{t}$. (ii)
Second, we note that the additional line broadening is universal,
i.e. independent to specific energy/mass of the line or the light
source or its components, so the broadening of line can be universally
attributed to the broadening or smearing of spacetime itself, which
is the requirement of an equivalence principle at the quantum level
\citep{2024arXiv240809630L}. (iii) Third, this additional broadening
effect is different from other line-broadening effects, e.g. the usual
thermal broadening (no matter the thermal temperature $T$ is from
the thermal motion or Unruh effect). Because the usual thermal broadening
$\delta\omega$ is particle mass $m$ dependent $\delta\omega\propto\omega_{0}\sqrt{\frac{T}{m}}$,
but the present effect $\delta\omega\propto a$ is universal. (iv)
And fourth, we also note that the additional broadening is related
to the squared $a$, so no matter it is an acceleration $+|a|$ or
deceleration $-|a|$, the spectrum is always broadened (not narrowed),
\begin{equation}
\left|a\right|=\sqrt{\frac{1}{2}\delta\sigma_{\omega}}.\label{eq:variance}
\end{equation}
That is to say that, the acceleration broadening is an irreversible
and history-dependent process of losing information or increasing
entropy, although not so completely like the Unruh effect.

The formula (\ref{eq:variance}) is written by the natural unit, in
the standard meter-kilogram-second unit (IS), we have $a=\frac{d^{2}x}{c^{2}dt^{2}}=\frac{1}{c^{2}}a_{IS}$,
and if one replaces the angular frequency $\omega=2\pi f$ by the
frequency $f$, it leads to
\begin{equation}
\frac{\left|a_{IS}\right|}{2\pi c^{2}}=\sqrt{\frac{1}{2}\delta\sigma_{f}}.\label{eq:variance_IS}
\end{equation}
Thus it is clearly that the broadening of the frequency $\sigma_{f}$
from the acceleration $a_{IS}$ is suppressed to be a very tiny effect,
because of the squared speed of light in the denominator.

Because both eq.(\ref{eq:accelerated frequency}) and eq.(\ref{eq:short time doppler shift})
are the Doppler shift, so in this sense, the Unruh effect and this
local short-time acceleration broadening effect both emerge as a consequence
of the Doppler shift of the accelerated wave in essential. In this
sense, the effect can be roughly understood as a local version of
Unruh effect. The differences are as follows. (a) First, because $\rho_{sho-acc}$
peaks at the identical center of the unaccelerated line, so it does
not completely lost the original frequency information of the line
during the short-time acceleration, it is only broadened or smeared;
but the information is completely lost in the long time acceleration,
in other words, the Unruh effect causes a line to blur into a continuous
Planckian spectrum. Obviously, the short-time acceleration spectrum
is not a Planckian thermalized type with maximal entropy, the state
of the short-time accelerated line is still almost a pure state, rather
than a complete mixed state (as the Unruh effect). (b) Second, as
is shown that the phase shift of the wave vector has higher order
$O(t^{3})$ than the one of frequency part, so the short-time acceleration-induced
distortion effect to the wave vector does exist, and dominated along
the spatial direction of the acceleration, although it is at higher
order than the frequency broadening. As a consequence, the short-time
acceleration-induced distortion of the wavefunction is spatially anisotropic
and local (taking low entropy), which is different from that the Unruh
effect induced thermal bath is isotropic and homogeneous (taking maximal
entropy).

\section{Possible Implications in Cosmology}

\subsection{Cosmic Accelerated Expansion induced Line Broadening }

As the Unruh effect, the short-time acceleration-induced line broadening
is also a tiny effect compared to other known broadening effects,
for example the thermal broadening or Doppler broadening etc, so the
effect is expected to be buried in many other broadening effects and
is ignorable in ordinary lab setup. There is an exception that significant
acceleration, and hence unignorable acceleration-induced broadening,
may appear. Note that in a spherical symmetric static deSitter metric,
the acceleration is proportional to the distance $r$ from a rest
earth observer, $a_{ds}=\frac{\Lambda}{3}r$, where $\Lambda$ the
cosmological constant, it seems there will be a large enough acceleration
at a cosmic scale distance $r$ and hence the line broadening would
expect unignorable at the cosmic scale.

Gauge equivalent to the spherical symmetric static deSitter metric,
we consider the accelerated expansion Friedmann-Robertson-Walker (FRW)
universe at the near current epoch or low redshift. The recession
velocity of a distant comoving tracer (e.g. stars or the type Ia supernovae)
w.r.t. an earth observer at redshift $z\lesssim O(1)$ is about $v\approx cz$
(in the unit $c=1$), so the related recession acceleration can be
given by \citep{Weinberg:2008zzc} 
\begin{equation}
a_{\Lambda}\approx\frac{1}{z}\frac{dz}{dt}\approx-q_{0}H_{0}\approx10^{-10}m/s^{2},\quad\left(z\lesssim O(1)\right),\label{eq:a0}
\end{equation}
where $H_{0}$ is the current Hubble's expansion rate, and $q_{0}\approx-0.64$
is the deceleration parameter. The acceleration is connected to the
familiar radial dependent acceleration in the static deSitter metric,
$a_{ds}=a_{\Lambda}z\approx\frac{\Lambda}{3}r$ by using $H_{0}r\approx z$
and $\Lambda\approx-3q_{0}H_{0}^{2}$, we will also discuss the relation
between these accelerations in different coordinates later.

For the relative low redshift $z\sim O(1)$ spectral lines, they are
comoving with the accelerated expanding space background and hence
broadened according to the proposed effect, (the Planckian cosmic
microwave background (CMB) spectrum is at the extremely high redshift
$z\approx1100$, whether it is due to the Unruh type acceleration
effect is beyond the scope of the paper). Therefore, besides the intrinsic
line width related to its lifetime (the first term of eq.(\ref{eq:full width})),
the second term, i.e. eq.(\ref{eq:a^2}), gives rise to an additional
universal line broadening coming from the cosmic acceleration expansion
\begin{equation}
\delta\sigma_{\omega}=2a_{\Lambda}^{2}\approx2q_{0}^{2}H_{0}^{2},\label{eq:background_width}
\end{equation}
which is isotropic and homogeneous as the background acceleration
$a_{\Lambda}$. Then it leads to a relative redshift broadening
\begin{equation}
\frac{\sigma_{z}}{z^{2}}\equiv\frac{\delta\sigma_{\omega}}{z^{2}\omega^{2}}\approx\frac{2q_{0}^{2}H_{0}^{2}}{\Delta\omega^{2}}.\label{eq:ratio}
\end{equation}
Since in our consideration, the line broadening $\delta\sigma_{\omega}$
is non-zero, so if the frequency shift $\Delta\omega$ is able to
tend to zero, the ratio becomes unphysically infinite. In practice,
at the cosmic scale, the period of a spectral line approaches to a
finite Hubble time, so the shift of its frequency should exist a lower
limit $\Delta\omega\rightarrow H_{0}$, in this case, the relative
redshift broadening tends to a constant of order $O(1)$
\begin{equation}
\frac{\sigma_{z}}{z^{2}}=\frac{\langle\delta z\rangle^{2}}{\langle z\rangle^{2}}\rightarrow2q_{0}^{2}\approx0.82,\label{eq:variance over z^2}
\end{equation}
when we take the measured value $q_{0}\approx-0.64$. Although it
is not exact, the ratio roughly agrees with the result from the Ricci
flow of quantum spacetime within errors \citep{Luo2014The,Luo2015Dark,Luo:2015pca,Luo:2019iby,Luo:2021zpi,Luo:2022goc,Luo:2022statistics,Luo:2022ywl,2023AnPhy.45869452L,2024chinarxiv,2024arXiv240809630L}.
The additional broadening is also independent to specific energy of
the line, so it universally reflects the property of the acceleration
of the frame or spacetime itself, which is an indication of the equivalence
principle at the quantum level. 

Compared to a standard distance-redshift relation without acceleration
$d_{L}=\frac{1}{H_{0}}\left(z+\frac{1}{2}z^{2}+...\right)$, where
$d_{L}$ is the luminosity distance between a supernova and a rest
earth observer. The acceleration of the expanding universe or the
``dark energy'' is encoded in the distance-redshift relation at
the quadratic order $O(z^{2})$, rather than directly be measured
from the redshift broadening or variance $\sigma_{z}$. Until now
the difficulty that we have not directly measured the line broadening
from the data of supernovae may be attributed to the fact that the
lines of supernovae at redshift $z\sim O(1)$ are too faint to fit
their profiles and extract their width or variance, but measuring
their central values to fit the distance-redshift relation to the
quadratic order is relatively more convenient and feasible. Since
the variance $\sigma_{z}$ does not modifies the linear term (central
value $\langle z\rangle$ of the redshift) but replacing the quadratic
term by $\langle z^{2}\rangle=\langle z\rangle^{2}+\langle\delta z\rangle^{2}=\langle z\rangle^{2}+\sigma_{z}$,
so we obtain the relation with a non-trivial acceleration 
\begin{equation}
d_{L}=\frac{1}{H_{0}}\left[\langle z\rangle+\frac{1}{2}\langle z^{2}\rangle+...\right]\approx\frac{1}{H_{0}}\left[\langle z\rangle+\frac{1}{2}\left(1+0.82\right)\langle z\rangle^{2}+...\right],
\end{equation}
although the deceleration parameter $-0.82$ only roughly agrees with the
fitting $-0.64$, the broadening effect in principle produces a modification
at the quadratic order, the unimportant discrepancy may come from
the estimate of $\Delta\omega$ in eq.(\ref{eq:ratio}). At the cosmic
scale, the cosmic acceleration induced redshift variance $\sigma_{z}\equiv\langle\delta z\rangle^{2}$
proportional to the mean squared redshift $\langle z\rangle^{2}$
cannot be ignored, especially at $\langle z\rangle\sim O(1)$. And
we can see that the measurement of the acceleration of the cosmic
expansion is essentially the measurement of the line broadening or
redshift variance $\sigma_{z}$. 

Although this interpretation of the cosmic acceleration seems like
a side-effect of spectral line, as the quantum version of equivalence
principle asserts \citep{2024arXiv240809630L}, the cosmic acceleration
can not be physically distinguished from the universal spectral line
broadening and/or spacetime coarse-graining, and hence it is physically
real. This is more or less similar with the fact that, due to the
universality of the cosmic redshift, the universal spectral line redshift
has been attributed to the expansion property of space, rather than
the property of the astronomical object's own motion, (only when the
object's proper motion is dominant at low-redshift, different objects
have different Doppler shifts, and hence they are not universal, so
that the redshifts represent their proper motion rather than the expansion
of the space). Similarly, due to the universality of the cosmic line
broadening (independent of spectral line energy), the cosmic line
broadening has also been attributed to the property of the accelerated
expansion of the space. The universal cosmic redshift of lines and
the expansion velocity of the space are equivalent and both real,
for the same reason, the universal cosmic broadening of lines and
the acceleration of the space are equivalent and both real, in the
sense of the quantum equivalence principle. 

The interpretation of the cosmic acceleration expansion also explains
the so-called ``coincidence problem'' as part of the cosmological
constant problem. Because no matter which epoch an observer lives
in the history of the universe, line broadening will always become
important in the distance-redshift relation at a near-current-redshift
w.r.t. the observer, so observer at any epoch of the history of universe
will see a ``recent start'' acceleration expansion. 

Because at the redshift $z\sim O(1)$, proper motion of an astronomical
object becomes small, so other line broadening effects coming from
the proper motion, such as thermal broadening or Doppler broadening,
become relatively small at $z\sim O(1)$. So if the effect does exist,
the universal cosmic acceleration induced line broadening effect may
gradually become dominant, and we expect that a directly measurement
of the variance (\ref{eq:variance over z^2}) may be possible in the
future, and if so, it would be an alternative indication of the accelerated
expansion or ``dark energy'', and also an evidence to the acceleration-induced
line broadening effect.

\subsection{Acceleration Discrepancy of Rotating Galaxy in Accelerated Expansion
Background}

The ``dark matter problem'' is a wide and complex issue unsolved,
including the rotation curve anomaly and the related acceleration
discrepancy at the outskirt of galaxies, missing matter problems in
the gravitational lensing (e.g. the \textquotedblleft bullet cluster\textquotedblright{}
being a famous example), and missing components in the CMB and the
cosmic evolution, etc. Here in this subsection, leaving the other
problems open, we only focus on the acceleration discrepancy problem,
since it possibly has a unified perspective with the cosmic accelerated
expansion, that is, it can also be readily understood (in part) by
the acceleration-induced line broadening effect.

Considering the radial acceleration at the radius $r$ to the center
of the galaxy, is given by $a_{D}=\frac{v_{R}^{2}}{r}$, the subscript
``$D$'' represents that it is given by the rotation velocity $v_{R}$
usually measured by the Doppler broadening of tracer (e.g. neutral
hydrogen that emits 21cm spectral lines), which comes from the red
and blue shift (rotation that away and towards) w.r.t. a rest earth
observer. It is worth stressing that the so-called Doppler broadening
coming from the proper rotational motion of galaxies should not be
confused with the proposed $a_{D}$-induced line broadening eq.(\ref{eq:variance}).
The so-called Doppler broadening is the major contribution to the
observed line profile from the outskirt of galaxies at low redshift
$z\ll1$, and it is much larger than $a_{D}$-induced line broadening
eq.(\ref{eq:variance}). As a consequence, what we measure, in the
Tully-Fisher relation, is the broadening coming from almost the so-called
Doppler broadening, rather than the acceleration broadening. This
is very different from the previous case of the type-Ia supernovae,
which are at $z\sim O(1)$ where their proper motions are relatively
small compared to their co-accelerating-expansion with the space,
so the line profile broadening is dominated by $a_{\Lambda}$-induced
broadening (and hence, in principle, it may be directly measured from,
e.g. supernovae, as we have proposed). 

In a usual Minkovski spacetime, $a_{D}$ certainly gives rise to all
the acceleration induced line broadening. However, it is not the case
in an accelerated expanding metric, in which a radial acceleration
coming from the proper rotation velocity, i.e. $a_{D}$, only produces
one part of the total acceleration induced line broadening $2a_{T}^{2}$,
subscript ``$T$'' is for an ``effective total acceleration''
given by the total acceleration induced line broadening, another part
inevitably comes form the background cosmic acceleration $a_{\Lambda}$.
For galaxies at low redshift $z\ll1$, their proper rotational motions
are dominant, so the $a_{\Lambda}$ is insignificant than $a_{D}$
in the total $a_{T}$ in this case. The role of $a_{\Lambda}$ correcting
$a_{D}$ is in a relative not-so-straightforward way, shown below. 

Let us consider the rotating galaxy is immersed in the accelerated
expansion background, so the total acceleration induced line broadening,
w.r.t. a rest earth observer, is not merely the part coming from the
proper rotation $2a_{D}^{2}$, but rather 
\begin{equation}
\delta\sigma_{\omega}=2a_{D}^{2}+2a_{\Lambda}^{2}=2a_{T}^{2}.\label{eq:total_width}
\end{equation}
It is the analogous version of the eq.(\ref{eq:background_width})
in the case of low redshift where proper rotational motion of galaxies
are also important, and $a_{\Lambda}^{2}\approx q_{0}^{2}H_{0}^{2}\approx\frac{q_{0}}{3}\Lambda$
(given by eq.(\ref{eq:a0})) is the cosmic acceleration of the background
coordinates w.r.t. the rest earth observer. As is claimed by the equivalence
principle \citep{2024arXiv240809630L}, the total broadening corresponding
to $a_{T}$ should be produced from all gravitational sources seen
by the rest earth observer, which includes not only baryonic matter
part ($a_{D}$-induced broadening) but also the vacuum ``dark energy''
part ($a_{\Lambda}$-induced broadening).

Note that the size of galaxy is also co-expanding with the space at
the expansion velocity $v_{E}=\frac{\dot{R}}{R}r=H_{0}r$, where $R(t)$
is the expanding scale factor of the Friedmann-Robertson-Walker (FRW)
universe, the overdot represents the FRW-time $t$ derivative, and
$r$ is the radial distance of a surrounding test body from the mass-center
of the galaxy. The rotation $v_{R}$ and expanding $v_{E}$ velocities
give extra kinetic energies $\frac{1}{2}\left(v_{R}^{2}+v_{E}^{2}\right)$
(in unit mass as the potential) to the galaxy, which effectively shifts
the static Newtonian potential by $\Phi_{N}\rightarrow\Phi_{N}+\frac{1}{2}\left(v_{R}^{2}+v_{E}^{2}\right)$.
As a consequence, the kinetic energies also give rise to extra contributions
to the total radial acceleration by the gradient operating on the
effective potential, 
\begin{equation}
a_{T}=\left|-\nabla_{r}\left(\Phi_{N}+\frac{1}{2}v_{E}^{2}+\frac{1}{2}v_{R}^{2}\right)\right|\overset{\textrm{local}}{=}\left|-\nabla_{r}\left(-\frac{GM_{b}}{r}+\frac{1}{2}\left(\frac{\dot{R}}{R}\right)^{2}r^{2}\right)\right|=\left|a_{N}+a_{E}\right|.\label{eq:aT=00003DaN+aE}
\end{equation}
where $\Phi_{N}=-\frac{GM_{b}}{r}$, $M_{b}$ is the baryonic mass.
In the formula, ``$\overset{\textrm{local}}{=}$'' stresses that
the acceleration is locally described in a static Schwarzschild coordinate
where the acceleration is defined by the gradient, distinguished from
``$\overset{\textrm{FRW}}{=}$'' in a FRW coordinate in the following
text. The rotation velocity $v_{R}$ is constant over the radius at
the outskirt of galaxy and hence is considered here gradient free
$\nabla_{r}v_{R}^{2}\approx0$. The $v_{E}$-gradient-induced acceleration
written in the local Schwarzschild coordinate is $a_{E}=\left|-\nabla_{r}\left(\frac{1}{2}v_{E}^{2}\right)\right|\overset{\textrm{local}}{=}\left|-\nabla_{r}\left(\frac{1}{2}\left(\frac{\dot{R}}{R}\right)^{2}r^{2}\right)\right|=\left(\frac{\dot{R}}{R}\right)^{2}r$,
the plus sign $+a_{E}$ in eq.(\ref{eq:aT=00003DaN+aE}) tells us
that it is pointing towards the mass-center of the galaxy corresponding
to an attractive force parallel with the Newtonian gravity force,
$a_{N}=\frac{GM_{b}}{r^{2}}$, and so it is in the opposite direction
of the repulsive force coming from the background cosmic acceleration
$a_{\Lambda}$ that is pointing away from the mass-center. When $r$
is of the size within galaxy, the effect of $a_{E}$ is relatively
small compared to the Newtonian gravity, because of the smallness
of the Hubble rate $\frac{\dot{R}}{R}\approx H_{0}$. But when $r$
is beyond the optical size of galaxies, say, to a critical radius
$r_{c}$ where $a_{E}\overset{\textrm{local}}{\approx}H_{0}^{2}r$
is comparable with $a_{N}=\frac{GM_{b}}{r^{2}}$, i.e. $r_{c}=\left(\frac{GM_{b}}{H_{0}^{2}}\right)^{1/3}$,
in this situation the influence of expanding-velocity induced attraction
$a_{E}$ to the dynamics of galaxies becomes unignorable and important. 

The ``acceleration'' in the ``local'' static Schwarzschild coordinate
and the surrounding ``FRW'' background are conceptually different,
the translation and crossover between these two context, more precisely,
the influence of the FRW expansion on the local dynamics are topics
of long history (see review \citep{Carrera:2008pi}). If considering
in a FRW time interval $dt$, the local space expands $v_{E}dt\overset{\textrm{FRW}}{=}Rdr+r\dot{R}dt$,
substituting $R=(1+z)^{-1}$ and $v_{E}=\frac{\dot{R}}{R}r$, so $dr=\left(R^{-1}-1\right)v_{E}dt\overset{\textrm{FRW}}{=}zv_{E}dt$,
see \citep{Carrera:2008pi}, and then using $v_{E}\overset{\textrm{FRW}}{\approx}z$,
we obtain the relation 
\begin{equation}
a_{E}=\left|-\nabla_{r}\left(\frac{1}{2}v_{E}^{2}\right)\right|=v_{E}\frac{dv_{E}}{dr}\overset{\textrm{FRW}}{\approx}\frac{1}{z}\frac{dz}{dt}\approx a_{\Lambda}.
\end{equation}
That is to say, when $a_{E}$ in a local Schwarzschild coordinate
is translated to a global FRW coordinate, the radial dependent $a_{E}$
observed by a galaxy's mass-center observer can be transformed to
a homogeneous and isotropic FRW background acceleration $a_{\Lambda}$
for observers at any places of the universe (including the earth).
So we replace $a_{E}$ by $a_{\Lambda}$ in the context of a FRW universe.
In this translation between these two context, we can also see that
the kinetic energies shifted local Schwarzschild space metric $g_{rr}\overset{\textrm{local}}{=}\left\{ 1+2\left[\Phi_{N}+\frac{1}{2}\left(v_{R}^{2}+v_{E}^{2}\right)\right]\right\} ^{-1}$
gradually crossovers to the FRW space metric, $g_{rr}\overset{\textrm{FRW}}{\longrightarrow}\left(1+v_{E}^{2}\right)^{-1}\overset{\textrm{FRW}}{\approx}\left(1+z^{2}\right)^{-1}\approx\left(1+z\right)^{-2}=R^{2}$
as the distance scale becomes larger and larger, say, $\Phi_{N}\overset{\textrm{FRW}}{\longrightarrow}0$
and the proper rotation motion gradually becomes much smaller than
the co-expansion, $v_{R}\overset{\textrm{FRW}}{\ll}v_{E}$.

On the other hand, it is also noted that, a body that is comoving
with the cosmological expansion is moving on an inertial trajectory
(i.e. force free), and by definition, the Newtonian dynamical force
is the cause for deviations from the inertial motion. So in the expansion
space background or the expansion size of the galaxy, the baryonic
Newtonian acceleration $a_{N}$ at the outskirt of a galaxy $r>r_{c}$
will not be the total $a_{T}$, but rather the relative part of $a_{T}$
exceeding the co-expanding ``local'' acceleration $a_{E}$, or equivalently,
the ``FRW'' background acceleration $a_{\Lambda}$ of the galaxy

\begin{equation}
a_{N}\overset{\textrm{FRW}}{=}a_{T}-a_{\Lambda}=\frac{GM_{b}}{r^{2}}.\label{eq:Newtonian gravity}
\end{equation}

To sum up, the discrepancy between the observed $a_{D}=\frac{v_{R}^{2}}{r}$
and the expected Newtonian $a_{N}=\frac{GM_{b}}{r^{2}}$ comes from
two aspects: on one hand, the background acceleration $a_{\Lambda}$
induced line broadening should be contributed to the total acceleration
$a_{T}$ induced line broadening, by eq.(\ref{eq:total_width}), which
is a new quantum effect coming from the spectral broadening by acceleration;
on the other hand, the background acceleration $a_{\Lambda}$ should
also be deducted from the total $a_{T}$, by eq.(\ref{eq:Newtonian gravity}),
so that Newton's gravity law applies in such a frame where the background
acceleration plays no role, which is a classical effect coming from
the influence of the cosmic expansion on the local galaxian dynamics.

\begin{figure}
\begin{centering}
\includegraphics[scale=0.5]{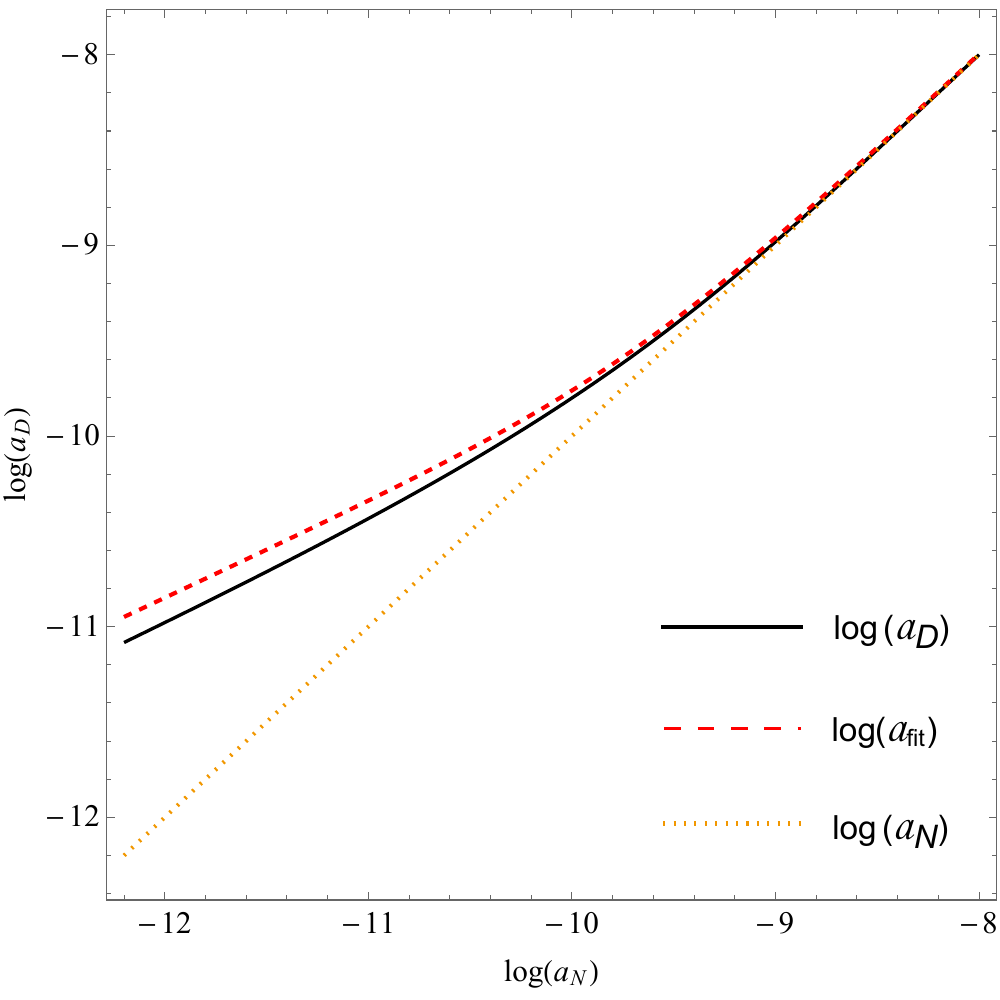}
\par\end{centering}
\caption{The relations eq.(\ref{eq:total_width}) and eq.(\ref{eq:Newtonian gravity})
give rise to a full acceleration discrepancy ($a_{D}\:\textrm{v.s.}\:a_{N}$
in unit $m/s^{2}$). The black solid curve is $a_{D}=\sqrt{\left(a_{N}+a_{0}\right)^{2}-a_{0}^{2}}$
in which $a_{\Lambda}$ in eq.(\ref{eq:interplation}) has been replaced
by $a_{0}\approx1.2\times10^{-10}m/s^{2}$ up to a factor of several
times. The curve is for any value of acceleration at any possible
radius of galaxies. $a_{D}$ agrees with the observation data points
(the red dashed curve is the data-points-fitting $a_{fit}=\frac{a_{N}}{1-e^{-\sqrt{a_{N}/a_{0}}}}$)
in \citep{Stacy2016Radial,2017ApJ...836..152L,Li:2018tdo} and deviates
from $a_{N}$ (the orange dotted line) at low acceleration where $a_{0}$
or $a_{\Lambda}$ becomes unignorable.}

\end{figure}
Finally, the radial acceleration corresponding to the proper rotation
velocity of a galaxy is given by 
\begin{equation}
a_{D}\overset{\textrm{FRW}}{=}\sqrt{\left(a_{N}+a_{\Lambda}\right)^{2}-a_{\Lambda}^{2}},\label{eq:interplation}
\end{equation}
in which we have combined the effects of the attractive $+a_{\Lambda}$
coming from eq.(\ref{eq:Newtonian gravity}) and the repulsive cosmological
constant or $-a_{\Lambda}^{2}$ coming from eq.(\ref{eq:total_width}).

The competition between the attractive $+a_{\Lambda}$ and repulsive
$-a_{\Lambda}^{2}$ can also be effectively viewed from the perspective
of a shifted Schwarzschild-deSitter (SdS) local metric \citep{Weinberg:2008zzc},
$g_{rr}^{(SdS)}=\left[1+2\Phi_{N}-\frac{1}{3}\Lambda r^{2}\right]^{-1}$
in which the repulsive effect (the repulsive potential $\frac{1}{6}\Lambda r^{2}$)
coming from the cosmological constant has been incorporated into the
Schwarzschild space metric $g_{rr}$. Considering the kinetic energies
shift $\Phi_{N}\rightarrow\Phi_{N}+\frac{1}{2}\left(v_{R}^{2}+v_{E}^{2}\right)$
as before, its radial metric becomes
\begin{equation}
g_{rr}^{(SdS)}\rightarrow\left[1+2\Phi_{N}-\frac{1}{3}\Lambda r^{2}+\left(v_{R}^{2}+v_{E}^{2}\right)\right]^{-1}=\left[1-\frac{2GM_{b}}{r}+\left(1-\Omega_{\Lambda}\right)H_{0}^{2}r^{2}+v_{R}^{2}\right]^{-1}.\label{eq:g00}
\end{equation}
Up to a gradient free term $v_{R}^{2}$, in the metric, the residual
attractive $\left(1-\Omega_{\Lambda}\right)H_{0}^{2}r^{2}>0$ potential
comes from the competition between the local attractive $v_{E}^{2}=H_{0}^{2}r^{2}$
and the repulsive $\frac{1}{3}\Lambda r^{2}$ terms. With appropriate
clumpy radial distribution assumption, e.g. $(1-\Omega_{\Lambda})H_{0}^{2}\propto\langle\rho_{M}\rangle\propto r^{-2}$,
it is this extra attractive term plays the effective role of the ``clumpy
missing matter'' which enhances the rotation acceleration, and hence
the rotation curve of test particles becomes constant at the outskirt
of galaxies.

We find that if the mutual competition between attractive and repulsive
terms is considered at the level of the metric (\ref{eq:g00}), a
distribution of the ``missing matter'' in a galaxy must be additionally
assumed so that the rotation curve is able to be fitted. However,
if the mutual competition is considered at the level of the acceleration
interpolation (\ref{eq:interplation}), there is no need for such
an additional assumption, because the ``distribution'' has already
implied in this interpolation function (\ref{eq:interplation}). This
is the beauty of the acceleration interpolation compared with the
``missing matter'' hypothesis.

The acceleration interpolating (\ref{eq:interplation}) also gives
rise to some empirical relations discussed at low $a_{N}$ limit.
Since $a_{\Lambda}$ is relatively a small constant acceleration compared
to $a_{D}$, so it is usually ignorable in ordinary lab setup, in
this sense $a_{N}$ and $a_{D}$ are approximately equal for $a_{D}\gg a_{\Lambda}$.
But when $a_{N}\rightarrow0$, for instance, at the outskirt of a
galaxy, so that the contribution of $a_{\Lambda}$ is unignorable.
In this situation, (\ref{eq:interplation}) reproduces MOND, see e.g.
\citep{Famaey:2011kh,Milgrom:2014usa}
\begin{equation}
a_{N}\approx\frac{a_{D}^{2}}{2a_{\Lambda}},\quad\left(a_{N}\rightarrow0\right).\label{eq:asymptotic eff acceleration}
\end{equation}
By using $a_{D}=\frac{v_{f}^{2}}{r}$ and $a_{N}=\frac{GM_{b}}{r^{2}}$,
where $v_{f}$ is the final constant rotation velocity at the outskirt
of galaxy, the relation straightforwardly gives the Baryonic-Tully-Fisher
relation $v_{f}^{4}\approx2GM_{b}a_{\Lambda}\sim GM_{b}a_{0}$ \citep{2000The},
where $a_{0}\approx1.2\times10^{-10}m/s^{2}$ is a fitting constant
parameter, up to a factor $\frac{a_{\Lambda}}{a_{0}}\approx(6\sim8)$.
So in this picture, the fitted constant $a_{0}$ of the Tully-Fisher
relation can be interpreted as the background cosmic acceleration
$a_{\Lambda}$ up to a factor. The difference between $a_{\Lambda}$
and $a_{0}$ can be attributed to various possibilities, e.g. because
the spiral galaxies are not point masses but rather flattened mass
distribution, and the geometric correction can be a few percent; and
there is also possibility that the residual undiscovered ``missing
matter'' does exist in part, in this sense, the correction brought
by the new acceleration effect only mitigates the ``missing matter
problem'' but does not exclude the dark matter completely, the theory
can be seen as a mixture of MOND-like at local and $\Lambda$CDM-like
at larger scale; and also possible that the renormalization of $a_{\Lambda}$
from the Ricci flow (through the renormalization of the background
curvature and the matter-to-gravity-coupling \citep{Luo:2022ywl},
etc. All these can be next leading order corrections in the framework.

At redshift $z\sim O(1)$, when proper motion of spectral lines (tracer)
becomes small, $a_{D}\rightarrow0$, so $a_{\Lambda}$ part becomes
more and more important, in this case, eq.(\ref{eq:total_width})
has a lower limit. Correspondingly $a_{T}$ appears to have a lower
bound, the bound is nothing but the background acceleration $a_{\Lambda}$
caused by the effective ``dark energy''. In usual spiral galaxies
at low redshift, the rotating part acceleration $a_{D}$ is able to
be distinguished from the background part $a_{\Lambda}$, because
of the evident rotation of the spiral galaxies. In this case, $a_{D}$
is unbounded from below, as is shown in Fig.1. However, for some faint
and non-spiral galaxies at relatively high redshift, in which the
random proper motion induced $a_{D}$ and cosmic background induced
$a_{\Lambda}$ are comparable, and distinguishing them becomes difficult
in observing the Doppler broadening of the lines, so that a mixed
up $a_{T}$ is the only measurable quantity. Consequently, a lower
bound of the mixed $a_{T}$ is possibly shown up in this situation
\citep{Luo:2022ywl}. Apart from the prediction (\ref{eq:variance over z^2}),
the possible lower bound in measuring a mixed $a_{T}$ for some faint
non-spiral galaxies is a second possible prediction of the effect.
Possible signs of the lower limit have also been indicated by some
data points of ultrafaint dwarf spheroidals \citep{2017ApJ...836..152L},
although the explanations are still controversial. 

\section{Conclusions}

In the paper, by a direct coordinate transforming to a local accelerated
frame and the Gabor transformation, we derive a quantum effect that
a non-uniform local short-time acceleration produces an additional
broadening to a spectral line, and the additional variance of the
spectral line is proportional to the square of the local acceleration
(\ref{eq:variance}), while the central value of the line remains
unaffected. The induced distortion effect to the spatial wave vector
of the spectrum is at higher order than the frequency broadening,
so the short-time acceleration effect is spatially anisotropic and
local. The effect can be considered as a generalization of the Unruh
effect. As we know that the velocity of a moving body can be measured
by the Doppler shift of the central value of the spectral line it
emits, according to the proposed effect, the acceleration of the moving
body can be measured by the broadening or variance of the line, and
the direction of the acceleration is given by the direction of the
wave vector distortion.

This effect provides us a possible alternative way to directly measure
the acceleration of the universe, which is equivalent to the standard
approach by fitting the deceleration parameter in the distance-redshift
relation. In short, the expansion rate of the universe is measured
by the universal redshift of lines, and the acceleration is by the
universal broadening of the lines. Since the effect in ordinary lab
setup is too small to be measured, (\ref{eq:variance over z^2}) being
of order one at the cosmic scale may be a feasible test of the prediction
in the future.

The cosmic acceleration broadening effect inevitably contributes to
the total acceleration induced line broadening emitted from a rotating
tracer around a galaxy, according to the composition relation (\ref{eq:total_width}).
Such contribution may lead to the radial acceleration discrepancy
of a rotating galaxy, which becomes significant especially at its
outskirt when the proper radial acceleration becomes small compared
to the cosmic background acceleration. The effect straightforwardly
leads to a MOND-like dynamics and the Tully-Fisher relation, so it
agrees with current observation data. The acceleration-induced line
broadening effect may provide us a simple and unified perspective
on the cosmic acceleration and the acceleration discrepancy of galaxies.

The measurement of the acceleration of the cosmic expansion (i.e.
the deceleration parameter $q_{0}$) in the distance-redshift relation
is, in essence, the measurement of the line broadening or redshift
variance $\sigma_{z}$. The additional line broadening from the background
cosmic acceleration also plays a key role in the acceleration discrepancy
at the outskirt of rotating galaxies. So we argue in the paper that
the additional acceleration effect on the line-broadening in cosmic
observations is crucial in understanding the missing components problems
of the universe at least in part.

More precisely, the concept of acceleration is merely approximate
and noncovariant, so the concept and effect proposed in the paper
can only be viewed as an intuitive and heuristic version of a more
fundamental and general covariant theory. It has been proposed in
the previous works that the eq.(\ref{eq:variance}) can be derived
from an underlying general coordinate transformation theory of quantum
version, namely a quantum reference frame and its renormalization
flow (a Ricci flow) developed by the author \citep{Luo2014The,Luo2015Dark,Luo:2015pca,Luo:2019iby}.
A quantum equivalence principle \citep{2024arXiv240809630L} guarantees
the equivalence between the acceleration induced universal line broadening
effect and the smearing or coarse-graining effect of the Ricci flow
of quantum spacetime. In the fundamental framework, not only is the
mean value of coordinate (classically) transformed, but also the second-order
moment quantum fluctuation or variance of coordinate is irreversibly
(quantum) broadening by the Ricci flow of the quantum spacetime, leading
to the diffeomorphism anomaly \citep{Luo:2021zpi} and an irreversible
H-theorem \citep{Luo:2022statistics} of the quantum spacetime, and
the cosmic acceleration expansion or the cosmological constant emerges
as the anomaly cancellation. The acceleration composition relation
(\ref{eq:total_width}) can be more fundamentally understood as a
relation that the curvature is modified by the cosmological constant
at leading order \citep{Luo:2022ywl,2024chinarxiv}. And the spectral
line broadening is essentially related to the broadening of kernel
of a Laplacian-like operator on a spacetime background, which is a
fundamental solution of the conjugate heat equation associated with
the Ricci flow of the quantum spacetime. This work provides an intuitive,
heuristic and alternative understanding of the key ideas and main
consequences of the mentioned fundamental theory. 
\begin{acknowledgments}
This work was supported in part by the National Science Foundation
of China (NSFC) under Grant No.11205149, and the Scientific Research
Foundation of Jiangsu University for Young Scholars under Grant No.15JDG153.

\bibliographystyle{plain}

\begin{thebibliography}{55}
\expandafter\ifx\csname natexlab\endcsname\relax\def\natexlab#1{#1}\fi
\expandafter\ifx\csname bibnamefont\endcsname\relax
  \def\bibnamefont#1{#1}\fi
\expandafter\ifx\csname bibfnamefont\endcsname\relax
  \def\bibfnamefont#1{#1}\fi
\expandafter\ifx\csname citenamefont\endcsname\relax
  \def\citenamefont#1{#1}\fi
\expandafter\ifx\csname url\endcsname\relax
  \def\url#1{\texttt{#1}}\fi
\expandafter\ifx\csname urlprefix\endcsname\relax\def\urlprefix{URL }\fi
\providecommand{\bibinfo}[2]{#2}
\providecommand{\eprint}[2][]{\url{#2}}

\bibitem[{\citenamefont{Stacy et~al.}(2016)\citenamefont{Stacy, S., McGaugh,
  Federico, Lelli, James, M., and Schombert}}]{Stacy2016Radial}
\bibinfo{author}{\bibnamefont{Stacy}}, \bibinfo{author}{\bibnamefont{S.}},
  \bibinfo{author}{\bibnamefont{McGaugh}},
  \bibinfo{author}{\bibnamefont{Federico}},
  \bibinfo{author}{\bibnamefont{Lelli}}, \bibinfo{author}{\bibnamefont{James}},
  \bibinfo{author}{\bibnamefont{M.}}, \bibnamefont{and}
  \bibinfo{author}{\bibnamefont{Schombert}}, \bibinfo{journal}{Physical Review
  Letters} \textbf{\bibinfo{volume}{117}}, \bibinfo{pages}{201101}
  (\bibinfo{year}{2016}).

\bibitem[{\citenamefont{{Lelli} et~al.}(2017)\citenamefont{{Lelli}, {McGaugh},
  {Schombert}, and {Pawlowski}}}]{2017ApJ...836..152L}
\bibinfo{author}{\bibfnamefont{F.}~\bibnamefont{{Lelli}}},
  \bibinfo{author}{\bibfnamefont{S.~S.} \bibnamefont{{McGaugh}}},
  \bibinfo{author}{\bibfnamefont{J.~M.} \bibnamefont{{Schombert}}},
  \bibnamefont{and} \bibinfo{author}{\bibfnamefont{M.~S.}
  \bibnamefont{{Pawlowski}}}, \bibinfo{journal}{\apj}
  \textbf{\bibinfo{volume}{836}}, \bibinfo{eid}{152} (\bibinfo{year}{2017}),
  \eprint{1610.08981}.

\bibitem[{\citenamefont{Li et~al.}(2018)\citenamefont{Li, Lelli, McGaugh, and
  Schombert}}]{Li:2018tdo}
\bibinfo{author}{\bibfnamefont{P.}~\bibnamefont{Li}},
  \bibinfo{author}{\bibfnamefont{F.}~\bibnamefont{Lelli}},
  \bibinfo{author}{\bibfnamefont{S.}~\bibnamefont{McGaugh}}, \bibnamefont{and}
  \bibinfo{author}{\bibfnamefont{J.}~\bibnamefont{Schombert}},
  \bibinfo{journal}{Astron. Astrophys.} \textbf{\bibinfo{volume}{615}},
  \bibinfo{pages}{A3} (\bibinfo{year}{2018}), \eprint{1803.00022}.

\bibitem[{\citenamefont{Unruh}(1976)}]{Unruh:1976db}
\bibinfo{author}{\bibfnamefont{W.~G.} \bibnamefont{Unruh}},
  \bibinfo{journal}{Phys. Rev. D} \textbf{\bibinfo{volume}{14}},
  \bibinfo{pages}{870} (\bibinfo{year}{1976}).

\bibitem[{\citenamefont{\v{S}oda et~al.}(2022)\citenamefont{\v{S}oda, Sudhir,
  and Kempf}}]{Soda:2021sql}
\bibinfo{author}{\bibfnamefont{B.}~\bibnamefont{\v{S}oda}},
  \bibinfo{author}{\bibfnamefont{V.}~\bibnamefont{Sudhir}}, \bibnamefont{and}
  \bibinfo{author}{\bibfnamefont{A.}~\bibnamefont{Kempf}},
  \bibinfo{journal}{Phys. Rev. Lett.} \textbf{\bibinfo{volume}{128}},
  \bibinfo{pages}{163603} (\bibinfo{year}{2022}), \eprint{2103.15838}.

\bibitem[{\citenamefont{Martin-Martinez
  et~al.}(2011)\citenamefont{Martin-Martinez, Fuentes, and
  Mann}}]{Martin-Martinez:2010gnz}
\bibinfo{author}{\bibfnamefont{E.}~\bibnamefont{Martin-Martinez}},
  \bibinfo{author}{\bibfnamefont{I.}~\bibnamefont{Fuentes}}, \bibnamefont{and}
  \bibinfo{author}{\bibfnamefont{R.~B.} \bibnamefont{Mann}},
  \bibinfo{journal}{Phys. Rev. Lett.} \textbf{\bibinfo{volume}{107}},
  \bibinfo{pages}{131301} (\bibinfo{year}{2011}), \eprint{1012.2208}.

\bibitem[{\citenamefont{Arya et~al.}(2022)\citenamefont{Arya, Mittal, Lochan,
  and Goyal}}]{Arya:2022lay}
\bibinfo{author}{\bibfnamefont{N.}~\bibnamefont{Arya}},
  \bibinfo{author}{\bibfnamefont{V.}~\bibnamefont{Mittal}},
  \bibinfo{author}{\bibfnamefont{K.}~\bibnamefont{Lochan}}, \bibnamefont{and}
  \bibinfo{author}{\bibfnamefont{S.~K.} \bibnamefont{Goyal}},
  \bibinfo{journal}{Phys. Rev. D} \textbf{\bibinfo{volume}{106}},
  \bibinfo{pages}{045011} (\bibinfo{year}{2022}), \eprint{2204.06595}.

\bibitem[{\citenamefont{Wadati and Miki}(2000)}]{Wadati2000The}
\bibinfo{author}{\bibnamefont{Wadati}} \bibnamefont{and}
  \bibinfo{author}{\bibnamefont{Miki}}, \bibinfo{journal}{J Phys Soc Jpn}
  \textbf{\bibinfo{volume}{68}}, \bibinfo{pages}{2543} (\bibinfo{year}{2000}).

\bibitem[{\citenamefont{Vandegrift and G.}(2000)}]{Vandegrift2000Accelerating}
\bibinfo{author}{\bibnamefont{Vandegrift}} \bibnamefont{and}
  \bibinfo{author}{\bibnamefont{G.}}, \bibinfo{journal}{American Journal of
  Physics} \textbf{\bibinfo{volume}{68}}, \bibinfo{pages}{576}
  (\bibinfo{year}{2000}).

\bibitem[{\citenamefont{Herdegen and Wawrzycki}(2002)}]{Herdegen:2001tt}
\bibinfo{author}{\bibfnamefont{A.}~\bibnamefont{Herdegen}} \bibnamefont{and}
  \bibinfo{author}{\bibfnamefont{J.}~\bibnamefont{Wawrzycki}},
  \bibinfo{journal}{Phys. Rev. D} \textbf{\bibinfo{volume}{66}},
  \bibinfo{pages}{044007} (\bibinfo{year}{2002}), \eprint{gr-qc/0110021}.

\bibitem[{\citenamefont{Huerfano et~al.}(2006)\citenamefont{Huerfano, Sahu, and
  Socolovsky}}]{2006Quantum}
\bibinfo{author}{\bibfnamefont{S.}~\bibnamefont{Huerfano}},
  \bibinfo{author}{\bibfnamefont{S.}~\bibnamefont{Sahu}}, \bibnamefont{and}
  \bibinfo{author}{\bibfnamefont{M.}~\bibnamefont{Socolovsky}},
  \bibinfo{journal}{Int.j.pure Appl.math} pp. \bibinfo{pages}{153--166}
  (\bibinfo{year}{2006}).

\bibitem[{\citenamefont{Matos}(2012)}]{2012Weak}
\bibinfo{author}{\bibfnamefont{C.~J.~D.} \bibnamefont{Matos}},
(\bibinfo{year}{2010}), \eprint{arXiv:1006:2657}.

\bibitem[{\citenamefont{Nauenberg and Michael}(2016)}]{Nauenberg2016Einstein}
\bibinfo{author}{\bibnamefont{Nauenberg}} \bibnamefont{and}
  \bibinfo{author}{\bibnamefont{Michael}}, \bibinfo{journal}{American Journal
  of Physics} \textbf{\bibinfo{volume}{84}}, \bibinfo{pages}{879}
  (\bibinfo{year}{2016}).

\bibitem[{\citenamefont{Seveso}(2017)}]{2017Does}
\bibinfo{author}{\bibfnamefont{L.}~\bibnamefont{Seveso}} \bibnamefont{and}
\bibinfo{author}{\bibfnamefont{V.}~\bibnamefont{Peri}} \bibnamefont{and}
\bibinfo{author}{\bibfnamefont{M.G.A}~\bibnamefont{Paris}}, \bibinfo{journal}{J. Phys. Conf. Ser.} \textbf{\bibinfo{volume}{880}}, \bibinfo{pages}{012067}
(\bibinfo{year}{2017}).

\bibitem[{\citenamefont{Colcelli et~al.}(2022)\citenamefont{Colcelli, Mussardo,
  Sierra, and Trombettoni}}]{2022Free}
\bibinfo{author}{\bibfnamefont{A.}~\bibnamefont{Colcelli}},
  \bibinfo{author}{\bibfnamefont{G.}~\bibnamefont{Mussardo}},
  \bibinfo{author}{\bibfnamefont{G.}~\bibnamefont{Sierra}}, \bibnamefont{and}
  \bibinfo{author}{\bibfnamefont{A.}~\bibnamefont{Trombettoni}},
  \bibinfo{journal}{American Journal of Physics} \textbf{\bibinfo{volume}{90}},
  \bibinfo{pages}{833} (\bibinfo{year}{2022}).

\bibitem[{\citenamefont{Emelyanov}(2022)}]{2022On}
\bibinfo{author}{\bibfnamefont{V.~A.} \bibnamefont{Emelyanov}},
  \bibinfo{journal}{The European Physical Journal C}
  \textbf{\bibinfo{volume}{82}},  (\bibinfo{year}{2022}).

\bibitem[{\citenamefont{Arya and Goyal}(2023)}]{Arya:2023okh}
\bibinfo{author}{\bibfnamefont{N.}~\bibnamefont{Arya}} \bibnamefont{and}
  \bibinfo{author}{\bibfnamefont{S.~K.} \bibnamefont{Goyal}},
  \bibinfo{journal}{Phys. Rev. D} \textbf{\bibinfo{volume}{108}},
  \bibinfo{pages}{085011} (\bibinfo{year}{2023}), \eprint{2305.19172}.

\bibitem[{\citenamefont{Alsing and Milburn}(2003)}]{Alsing:2003es}
\bibinfo{author}{\bibfnamefont{P.~M.} \bibnamefont{Alsing}} \bibnamefont{and}
  \bibinfo{author}{\bibfnamefont{G.~J.} \bibnamefont{Milburn}},
  \bibinfo{journal}{Phys. Rev. Lett.} \textbf{\bibinfo{volume}{91}},
  \bibinfo{pages}{180404} (\bibinfo{year}{2003}), \eprint{quant-ph/0302179}.

\bibitem[{\citenamefont{Alsing et~al.}(2006)\citenamefont{Alsing,
  Fuentes-Schuller, Mann, and Tessier}}]{Alsing:2006cj}
\bibinfo{author}{\bibfnamefont{P.~M.} \bibnamefont{Alsing}},
  \bibinfo{author}{\bibfnamefont{I.}~\bibnamefont{Fuentes-Schuller}},
  \bibinfo{author}{\bibfnamefont{R.~B.} \bibnamefont{Mann}}, \bibnamefont{and}
  \bibinfo{author}{\bibfnamefont{T.~E.} \bibnamefont{Tessier}},
  \bibinfo{journal}{Phys. Rev. A} \textbf{\bibinfo{volume}{74}},
  \bibinfo{pages}{032326} (\bibinfo{year}{2006}), \eprint{quant-ph/0603269}.

\bibitem[{\citenamefont{Bruschi et~al.}(2012)\citenamefont{Bruschi, Fuentes,
  and Louko}}]{Bruschi:2011ug}
\bibinfo{author}{\bibfnamefont{D.~E.} \bibnamefont{Bruschi}},
  \bibinfo{author}{\bibfnamefont{I.}~\bibnamefont{Fuentes}}, \bibnamefont{and}
  \bibinfo{author}{\bibfnamefont{J.}~\bibnamefont{Louko}},
  \bibinfo{journal}{Phys. Rev. D} \textbf{\bibinfo{volume}{85}},
  \bibinfo{pages}{061701} (\bibinfo{year}{2012}), \eprint{1105.1875}.

\bibitem[{\citenamefont{Adesso et~al.}(2012)\citenamefont{Adesso, Ragy, and
  Girolami}}]{Adesso:2012gw}
\bibinfo{author}{\bibfnamefont{G.}~\bibnamefont{Adesso}},
  \bibinfo{author}{\bibfnamefont{S.}~\bibnamefont{Ragy}}, \bibnamefont{and}
  \bibinfo{author}{\bibfnamefont{D.}~\bibnamefont{Girolami}},
  \bibinfo{journal}{Class. Quant. Grav.} \textbf{\bibinfo{volume}{29}},
  \bibinfo{pages}{224002} (\bibinfo{year}{2012}), \eprint{1205.0222}.

\bibitem[{\citenamefont{Friis et~al.}(2012)\citenamefont{Friis, Bruschi, Louko,
  and Fuentes}}]{Friis:2012tb}
\bibinfo{author}{\bibfnamefont{N.}~\bibnamefont{Friis}},
  \bibinfo{author}{\bibfnamefont{D.~E.} \bibnamefont{Bruschi}},
  \bibinfo{author}{\bibfnamefont{J.}~\bibnamefont{Louko}}, \bibnamefont{and}
  \bibinfo{author}{\bibfnamefont{I.}~\bibnamefont{Fuentes}},
  \bibinfo{journal}{Phys. Rev. D} \textbf{\bibinfo{volume}{85}},
  \bibinfo{pages}{081701} (\bibinfo{year}{2012}), \eprint{1201.0549}.

\bibitem[{\citenamefont{Toro\v{s} et~al.}(2020)\citenamefont{Toro\v{s},
  Restuccia, Gibson, Cromb, Ulbricht, Padgett, and Faccio}}]{Toros:2019ptw}
\bibinfo{author}{\bibfnamefont{M.}~\bibnamefont{Toro\v{s}}},
  \bibinfo{author}{\bibfnamefont{S.}~\bibnamefont{Restuccia}},
  \bibinfo{author}{\bibfnamefont{G.~M.} \bibnamefont{Gibson}},
  \bibinfo{author}{\bibfnamefont{M.}~\bibnamefont{Cromb}},
  \bibinfo{author}{\bibfnamefont{H.}~\bibnamefont{Ulbricht}},
  \bibinfo{author}{\bibfnamefont{M.}~\bibnamefont{Padgett}}, \bibnamefont{and}
  \bibinfo{author}{\bibfnamefont{D.}~\bibnamefont{Faccio}},
  \bibinfo{journal}{Phys. Rev. A} \textbf{\bibinfo{volume}{101}},
  \bibinfo{pages}{043837} (\bibinfo{year}{2020}), \eprint{1911.06007}.

\bibitem[{\citenamefont{Restuccia et~al.}(2019)\citenamefont{Restuccia,
  Toro\v{s}, Gibson, Ulbricht, Faccio, and Padgett}}]{Restuccia:2019chf}
\bibinfo{author}{\bibfnamefont{S.}~\bibnamefont{Restuccia}},
  \bibinfo{author}{\bibfnamefont{M.}~\bibnamefont{Toro\v{s}}},
  \bibinfo{author}{\bibfnamefont{G.~M.} \bibnamefont{Gibson}},
  \bibinfo{author}{\bibfnamefont{H.}~\bibnamefont{Ulbricht}},
  \bibinfo{author}{\bibfnamefont{D.}~\bibnamefont{Faccio}}, \bibnamefont{and}
  \bibinfo{author}{\bibfnamefont{M.~J.} \bibnamefont{Padgett}},
  \bibinfo{journal}{Phys. Rev. Lett.} \textbf{\bibinfo{volume}{123}},
  \bibinfo{pages}{110401} (\bibinfo{year}{2019}), \eprint{1906.03400}.

\bibitem[{\citenamefont{Toro\v{s} et~al.}(2022)\citenamefont{Toro\v{s}, Cromb,
  Paternostro, and Faccio}}]{Toros:2022tpq}
\bibinfo{author}{\bibfnamefont{M.}~\bibnamefont{Toro\v{s}}},
  \bibinfo{author}{\bibfnamefont{M.}~\bibnamefont{Cromb}},
  \bibinfo{author}{\bibfnamefont{M.}~\bibnamefont{Paternostro}},
  \bibnamefont{and} \bibinfo{author}{\bibfnamefont{D.}~\bibnamefont{Faccio}},
  \bibinfo{journal}{Phys. Rev. Lett.} \textbf{\bibinfo{volume}{129}},
  \bibinfo{pages}{260401} (\bibinfo{year}{2022}), \eprint{2207.14371}.

\bibitem[{\citenamefont{Rogers}(1988)}]{Rogers:1988zz}
\bibinfo{author}{\bibfnamefont{J.}~\bibnamefont{Rogers}},
  \bibinfo{journal}{Phys. Rev. Lett.} \textbf{\bibinfo{volume}{61}},
  \bibinfo{pages}{2113} (\bibinfo{year}{1988}).

\bibitem[{\citenamefont{Chen and Tajima}(1999)}]{Chen:1998kp}
\bibinfo{author}{\bibfnamefont{P.}~\bibnamefont{Chen}} \bibnamefont{and}
  \bibinfo{author}{\bibfnamefont{T.}~\bibnamefont{Tajima}},
  \bibinfo{journal}{Phys. Rev. Lett.} \textbf{\bibinfo{volume}{83}},
  \bibinfo{pages}{256} (\bibinfo{year}{1999}).

\bibitem[{\citenamefont{Lynch et~al.}(2021)\citenamefont{Lynch, Cohen, Hadad,
  and Kaminer}}]{Lynch:2019hmk}
\bibinfo{author}{\bibfnamefont{M.~H.} \bibnamefont{Lynch}},
  \bibinfo{author}{\bibfnamefont{E.}~\bibnamefont{Cohen}},
  \bibinfo{author}{\bibfnamefont{Y.}~\bibnamefont{Hadad}}, \bibnamefont{and}
  \bibinfo{author}{\bibfnamefont{I.}~\bibnamefont{Kaminer}},
  \bibinfo{journal}{Phys. Rev. D} \textbf{\bibinfo{volume}{104}},
  \bibinfo{pages}{025015} (\bibinfo{year}{2021}), \eprint{1903.00043}.

\bibitem[{\citenamefont{Lynch}(2022)}]{Lynch:2022nls}
\bibinfo{author}{\bibfnamefont{M.~H.} \bibnamefont{Lynch}}
(\bibinfo{year}{2022}), \eprint{arXiv:2205.06591}.

\bibitem[{\citenamefont{Famaey and McGaugh}(2012)}]{Famaey:2011kh}
\bibinfo{author}{\bibfnamefont{B.}~\bibnamefont{Famaey}} \bibnamefont{and}
  \bibinfo{author}{\bibfnamefont{S.}~\bibnamefont{McGaugh}},
  \bibinfo{journal}{Living Rev. Rel.} \textbf{\bibinfo{volume}{15}},
  \bibinfo{pages}{10} (\bibinfo{year}{2012}), \eprint{1112.3960}.

\bibitem[{\citenamefont{Milgrom}(2015)}]{Milgrom:2014usa}
\bibinfo{author}{\bibfnamefont{M.}~\bibnamefont{Milgrom}},
  \bibinfo{journal}{Can. J. Phys.} \textbf{\bibinfo{volume}{93}},
  \bibinfo{pages}{107} (\bibinfo{year}{2015}), \eprint{1404.7661}.

\bibitem[{\citenamefont{Milgrom}(1999)}]{Milgrom:1998sy}
\bibinfo{author}{\bibfnamefont{M.}~\bibnamefont{Milgrom}},
  \bibinfo{journal}{Phys. Lett. A} \textbf{\bibinfo{volume}{253}},
  \bibinfo{pages}{273} (\bibinfo{year}{1999}), \eprint{astro-ph/9805346}.

\bibitem[{\citenamefont{Verlinde}(2017)}]{Verlinde:2016toy}
\bibinfo{author}{\bibfnamefont{E.~P.} \bibnamefont{Verlinde}},
  \bibinfo{journal}{SciPost Phys.} \textbf{\bibinfo{volume}{2}},
  \bibinfo{pages}{016} (\bibinfo{year}{2017}), \eprint{1611.02269}.

\bibitem[{\citenamefont{Pikhitsa}(2010)}]{Pikhitsa:2010nd}
\bibinfo{author}{\bibfnamefont{P.~V.} \bibnamefont{Pikhitsa}}
(\bibinfo{year}{2010}), \eprint{arXiv:1010.0318}.

\bibitem[{\citenamefont{Klinkhamer and Kopp}(2011)}]{Klinkhamer:2011un}
\bibinfo{author}{\bibfnamefont{F.~R.} \bibnamefont{Klinkhamer}}
  \bibnamefont{and} \bibinfo{author}{\bibfnamefont{M.}~\bibnamefont{Kopp}},
  \bibinfo{journal}{Mod. Phys. Lett. A} \textbf{\bibinfo{volume}{26}},
  \bibinfo{pages}{2783} (\bibinfo{year}{2011}), \eprint{1104.2022}.

\bibitem[{\citenamefont{Crispino et~al.}(2008)\citenamefont{Crispino, Higuchi,
  and Matsas}}]{Crispino:2007eb}
\bibinfo{author}{\bibfnamefont{L.~C.~B.} \bibnamefont{Crispino}},
  \bibinfo{author}{\bibfnamefont{A.}~\bibnamefont{Higuchi}}, \bibnamefont{and}
  \bibinfo{author}{\bibfnamefont{G.~E.~A.} \bibnamefont{Matsas}},
  \bibinfo{journal}{Rev. Mod. Phys.} \textbf{\bibinfo{volume}{80}},
  \bibinfo{pages}{787} (\bibinfo{year}{2008}), \eprint{0710.5373}.

\bibitem[{\citenamefont{{Alsing} and {Milonni}}(2004)}]{2004AmJPh..72.1524A}
\bibinfo{author}{\bibfnamefont{P.~M.} \bibnamefont{{Alsing}}} \bibnamefont{and}
  \bibinfo{author}{\bibfnamefont{P.~W.} \bibnamefont{{Milonni}}},
  \bibinfo{journal}{American Journal of Physics} \textbf{\bibinfo{volume}{72}},
  \bibinfo{pages}{1524} (\bibinfo{year}{2004}), \eprint{quant-ph/0401170}.

\bibitem[{\citenamefont{Eliezer and C.}(1977)}]{Eliezer1977The}
\bibinfo{author}{\bibnamefont{Eliezer}} \bibnamefont{and}
  \bibinfo{author}{\bibfnamefont{J.}~\bibnamefont{C.}},
  \bibinfo{journal}{American Journal of Physics} \textbf{\bibinfo{volume}{45}},
  \bibinfo{pages}{1218} (\bibinfo{year}{1977}).

\bibitem[{\citenamefont{Daubechies and Heil}(1992)}]{1992Ten}
\bibinfo{author}{\bibfnamefont{I.}~\bibnamefont{Daubechies}} \bibnamefont{and}
  \bibinfo{author}{\bibfnamefont{C.}~\bibnamefont{Heil}},
  \bibinfo{journal}{Computers in Physics} \textbf{\bibinfo{volume}{6}},
  \bibinfo{pages}{697} (\bibinfo{year}{1992}).

\bibitem[{\citenamefont{Colella et~al.}(1975)\citenamefont{Colella, Overhauser,
  and Werner}}]{Colella:1975dq}
\bibinfo{author}{\bibfnamefont{R.}~\bibnamefont{Colella}},
  \bibinfo{author}{\bibfnamefont{A.~W.} \bibnamefont{Overhauser}},
  \bibnamefont{and} \bibinfo{author}{\bibfnamefont{S.~A.}
  \bibnamefont{Werner}}, \bibinfo{journal}{Phys. Rev. Lett.}
  \textbf{\bibinfo{volume}{34}}, \bibinfo{pages}{1472} (\bibinfo{year}{1975}).

\bibitem[{\citenamefont{Peters et~al.}(1999)\citenamefont{Peters, Chung, and
  Chu}}]{1999Measurement}
\bibinfo{author}{\bibfnamefont{A.}~\bibnamefont{Peters}},
  \bibinfo{author}{\bibfnamefont{K.~Y.} \bibnamefont{Chung}}, \bibnamefont{and}
  \bibinfo{author}{\bibfnamefont{S.}~\bibnamefont{Chu}},
  \bibinfo{journal}{Nature} \textbf{\bibinfo{volume}{400}},
  \bibinfo{pages}{849} (\bibinfo{year}{1999}).

\bibitem[{\citenamefont{{Luo}}(2024{\natexlab{a}})}]{2024arXiv240809630L}
\bibinfo{author}{\bibfnamefont{M.~J.} \bibnamefont{{Luo}}},
(\bibinfo{year}{2024}{\natexlab{a}}), \eprint{arXiv:2408.09630}.

\bibitem[{\citenamefont{Weinberg}(2008)}]{Weinberg:2008zzc}
\bibinfo{author}{\bibfnamefont{S.}~\bibnamefont{Weinberg}},
  \emph{\bibinfo{title}{{Cosmology}}} (\bibinfo{year}{2008}), ISBN
  \bibinfo{isbn}{978-0-19-852682-7}.

\bibitem[{\citenamefont{Luo}(2014)}]{Luo2014The}
\bibinfo{author}{\bibfnamefont{M.~J.} \bibnamefont{Luo}},
  \bibinfo{journal}{Nuclear Physics} \textbf{\bibinfo{volume}{884}},
  \bibinfo{pages}{344} (\bibinfo{year}{2014}).

\bibitem[{\citenamefont{Luo}(2015)}]{Luo2015Dark}
\bibinfo{author}{\bibfnamefont{M.~J.} \bibnamefont{Luo}},
  \bibinfo{journal}{Journal of High Energy Physics}
  \textbf{\bibinfo{volume}{06}}, \bibinfo{pages}{1} (\bibinfo{year}{2015}).

\bibitem[{\citenamefont{Luo}(2018)}]{Luo:2015pca}
\bibinfo{author}{\bibfnamefont{M.~J.} \bibnamefont{Luo}},
  \bibinfo{journal}{Int. J. Mod. Phys.} \textbf{\bibinfo{volume}{D27}},
  \bibinfo{pages}{1850081} (\bibinfo{year}{2018}), \eprint{1507.08755}.

\bibitem[{\citenamefont{Luo}(2021{\natexlab{a}})}]{Luo:2019iby}
\bibinfo{author}{\bibfnamefont{M.~J.} \bibnamefont{Luo}},
  \bibinfo{journal}{Found. Phys.} \textbf{\bibinfo{volume}{51}},
  \bibinfo{pages}{2} (\bibinfo{year}{2021}{\natexlab{a}}), \eprint{1907.05217}.

\bibitem[{\citenamefont{Luo}(2021{\natexlab{b}})}]{Luo:2021zpi}
\bibinfo{author}{\bibfnamefont{M.~J.} \bibnamefont{Luo}},
  \bibinfo{journal}{Class. Quant. Grav.} \textbf{\bibinfo{volume}{38}},
  \bibinfo{pages}{155018} (\bibinfo{year}{2021}{\natexlab{b}}),
  \eprint{2106.16150}.

\bibitem[{\citenamefont{Luo}(2022)}]{Luo:2022goc}
\bibinfo{author}{\bibfnamefont{M.~J.} \bibnamefont{Luo}},
  \bibinfo{journal}{Annals Phys.} \textbf{\bibinfo{volume}{441}},
  \bibinfo{pages}{168861} (\bibinfo{year}{2022}), \eprint{2201.10732}.

\bibitem[{\citenamefont{Luo}(2023{\natexlab{a}})}]{Luo:2022statistics}
\bibinfo{author}{\bibfnamefont{M.~J.} \bibnamefont{Luo}},
  \bibinfo{journal}{Int. J. Mod. Phys. D} \textbf{\bibinfo{volume}{32}},
  \bibinfo{pages}{2350022} (\bibinfo{year}{2023}{\natexlab{a}}),
  \eprint{2302.08651}.

\bibitem[{\citenamefont{Luo}(2023{\natexlab{b}})}]{Luo:2022ywl}
\bibinfo{author}{\bibfnamefont{M.~J.} \bibnamefont{Luo}},
  \bibinfo{journal}{Int. J. Theor. Phys.} \textbf{\bibinfo{volume}{62}},
  \bibinfo{pages}{91} (\bibinfo{year}{2023}{\natexlab{b}}),
  \eprint{2210.06082}.

\bibitem[{\citenamefont{{Luo}}(2023)}]{2023AnPhy.45869452L}
\bibinfo{author}{\bibfnamefont{M.~J.} \bibnamefont{{Luo}}},
  \bibinfo{journal}{Annals of Physics} \textbf{\bibinfo{volume}{458}},
  \bibinfo{eid}{169452} (\bibinfo{year}{2023}), \eprint{2112.00218}.

\bibitem[{\citenamefont{{Luo}}(2024{\natexlab{b}})}]{2024chinarxiv}
\bibinfo{author}{\bibfnamefont{M.~J.} \bibnamefont{{Luo}}},
  \bibinfo{journal}{ChinarXiv}  (\bibinfo{year}{2024}{\natexlab{b}}),
  \eprint{202404.00156 (in Chinese)}.

\bibitem[{\citenamefont{Carrera and Giulini}(2010)}]{Carrera:2008pi}
\bibinfo{author}{\bibfnamefont{M.}~\bibnamefont{Carrera}} \bibnamefont{and}
  \bibinfo{author}{\bibfnamefont{D.}~\bibnamefont{Giulini}},
  \bibinfo{journal}{Rev. Mod. Phys.} \textbf{\bibinfo{volume}{82}},
  \bibinfo{pages}{169} (\bibinfo{year}{2010}), \eprint{0810.2712}.

\bibitem[{\citenamefont{Mcgaugh et~al.}(2000)\citenamefont{Mcgaugh, Schombert,
  Bothun, and Blok}}]{2000The}
\bibinfo{author}{\bibfnamefont{S.~S.} \bibnamefont{Mcgaugh}},
  \bibinfo{author}{\bibfnamefont{J.~M.} \bibnamefont{Schombert}},
  \bibinfo{author}{\bibfnamefont{G.~D.} \bibnamefont{Bothun}},
  \bibnamefont{and} \bibinfo{author}{\bibfnamefont{W.~D.} \bibnamefont{Blok}},
  \bibinfo{journal}{Astrophysical Journal}  (\bibinfo{year}{2000}).

\end{thebibliography}

\end{acknowledgments}

\end{document}